# High temperature decomposition and age hardening of single-phase wurtzite Ti$_{1-x}$Al$_x$N thin films grown by cathodic arc deposition

J. Salamania[1,2*], F. Bock[3], L.J.S. Johnson[4], F. Tasnádi[3], K.M. Calamba Kwick[4], A.F. Farhadizaeh[1], I.A. Abrikosov[3], L. Rogström[1], M. Odén[1]

[1]*Nanostructured Materials Division, Department of Physics, Chemistry and Biology (IFM), Linköping University, Linköping, SE-581 83, Sweden*
[2]*Seco Tools AB, Fagersta, SE-737 82, Sweden*
[3]*Theoretical Physics Division, Department of Physics, Chemistry and Biology (IFM), Linköping University, Linköping, SE-581 83, Sweden*
[4]*Sandvik Coromant AB, Stockholm, SE-126 79, Sweden*

*corresponding author: janella.salamania@liu.se

**Abstract**

We investigated the high temperature decomposition behavior of wurtzite phase Ti$_{1-x}$Al$_x$N films using experimental methods and first-principles calculations. Single phase metastable wurtzite Ti$_{1-x}$Al$_x$N (x = 0.65, 0.75, 085 and 0.95) solid solution films were grown by cathodic arc deposition using low duty cycle pulsed substrate-bias voltage. First-principles calculated elastic constants of the wurtzite Ti$_{1-x}$Al$_x$N phase show a strong dependence on alloy composition. The predicted phase diagram shows a miscibility gap with an unstable region. High resolution scanning transmission electron microscopy and chemical mapping demonstrate decomposition of the films after high temperature annealing (950°C), which resulted in nanoscale chemical compositional modulations containing Ti-rich and Al-rich regions with coherent or semi coherent interfaces. This spinodal decomposition of the wurtzite film causes age hardening of 1-2 GPa.

1. Introduction

For many years, titanium aluminum nitride (TiAlN) has been an established and reliable material system for hard coating applications [1]. Besides being a hard material, its attractiveness stems from its ability to undergo spinodal decomposition (of its cubic solid solution) into coherent c-TiN-rich and c-AlN-rich domains [2,3] and results in age hardening when heated [4], which is beneficial when used as a protective coating on metal cutting tools [5]. However, TiAlN films also typically display a second phase transformation of the c-AlN rich domains into their more stable wurtzite w-AlN form during decomposition [6,7]. Studies of high aluminum-containing $Ti_{1-x}Al_xN$ films ($x \geq 0.70$) grown by physical vapor deposition (PVD) techniques are thus scarce and traditionally neglected due to the prevailing notion that the presence of the wurtzite phase deteriorates the mechanical performance of TiAlN-based hard coatings [2,8,9]. The worsened wear properties of coatings with high Al content are attributed to the softer nature of the formed wurtzite AlN phase and strain changes [4]. Because of this, research has been focused on cubic TiAlN and to suppress the formation of the unwanted wurtzite TiAlN phase by changing the alloy composition [10-12], defect population [13-16], and microstructure [17-19]. The strong focus on the cubic phase also resulted in a lack of information on how the wurtzite phase behaves at elevated temperatures.

Several reports, however, claim that w-AlN rich domains that are semicoherent with c-TiAlN domains can positively affect mechanical properties at higher temperatures [17,20,21]. *Ab-initio* studies have also shown that c-$Ti_{0.25}Al_{0.75}N$ solid solutions exhibit desired room-temperature strength but significantly higher toughness than c-TiN and c-$Ti_{0.5}Al_{0.5}N$ owing to its activation of local lattice transformations to its more stable B4 wurtzite polymorph that dissipates stress [22]. Single phase wurtzite coatings were also found to have the best oxidation resistance when compared to cubic and mixed cubic and wurtzite $Ti_{1-x}Al_xN$ coatings [23]. Such findings indicate that the wurtzite phase is not just detrimental and instead could be used to engineer improved TiAlN coatings.

Another hindrance to understanding wurtzite TiAlN is the difficulty of synthesizing it as single-phase solid solution, although some attempts exist [2,24-26]. One of the earliest attempts was by Hörling *et al.* who used cathodic arc evaporation that resulted in a $Ti_{0.26}Al_{0.74}N$ coating with nanocrystallites of w-AlN embedded in a cubic c-TiAlN matrix [2]. Shimizu *et al.* have shown that highly textured wurtzite TiAlN phases (x = 0.70-0.73) can be grown at low temperatures by high power impulse magnetron sputtering (HiPIMS) [24]. The deposited coatings exhibited excellent hardness (up to 35 GPa) despite the dominant presence of the wurtzite phase [24]. More recently, Calamba *et al.* [25] grew metastable wurtzite $Ti_{1-x}Al_xN$ (x = 0.77) films by an epitaxial stabilization technique by ultrahigh vacuum (UHV) magnetron sputtering, wherein the as-deposited wurtzite phase already showed a segregation into Al and Ti-rich domains. The microstructure remained stable after annealing at 900°C, suggesting better thermal stability of the wurtzite phase against coarsening [25]. Considering its known presence in industrial-grade nitride coatings [7], the thermal behavior of wurtzite phase merits further understanding.

Recently, we reported the possibility to deposit low-defect density films at high $N_2$ pressures by applying low duty-cycle pulsed bias voltage to the substrate [27]. By doing this, the combination of high $N_2$ pressures and short bias voltage on-times reduces the average energy supplied to the growth front, thereby reducing lattice defects and suppressing the number of sites for re-nucleation [27]. With this setup, it is possible to keep the composition ratio of the cathode in the film (especially important when growing metastable high Al-films) while limiting diffusion processes in the film such that pseudo-epitaxial growth is promoted with the cathodic arc technique.

In this study, the aim is to fill the knowledge gap in the high Al regime, *i.e.*, the wurtzite region of the TiAlN system. Theoretical studies and simulations of the thermal stability and subsequent spinodal decomposition of TiAlN are only focused on the metastable cubic phases [28-30] or its possible transition to the wurtzite phases [31,32]. Thus, little is known about the thermal stability and decomposition routes of a wurtzite $Ti_{1-x}Al_xN$ solid solution. Here, we combine high-resolution HAADF-STEM imaging and first-principles calculation to investigate the high-temperature decomposition of wurtzite $Ti_{1-x}Al_xN$ thin films.

2. Experimental Details

*2.1 Substrate and TiN Seed Layer Preparation*

Wurtzite $Ti_{1-x}Al_xN$ films weres deposited onto previously grown 85-120 nm-thick TiN(111) seed layers on MgO(111) and $Al_2O_3$(0001) c-cut substrates (10x10x0.5 $mm^3$ of >99.5 % purity). These single crystal TiN thin film seed layers were deposited using a Mantis ultrahigh vacuum DC magnetron sputtering system with a base pressure of less than 8 x $10^{-10}$ Torr (1 x $10^{-7}$ Pa) [33]. Prior to the deposition of TiN seed layers, the substrates were subjected to systematic chemical pre-cleaning with trichloroethylene, acetone, and ethanol [34], plus a 30 min vacuum anneal at 800°C in the deposition chamber. The TiN seed layers were deposited at a temperature of 800°C using a single element Ti target (99.9% purity) in a 20 sccm argon to 4 sccm nitrogen gas mixture, resulting in a working pressure of 0.3 Pa. More details of the TiN deposition is outlined in Ref. [33]. The TiN seed layer was deposited as a barrier to diffusion of Al into the MgO substrate and to prevent $MgAl_2O_4$ spinel formation [35,36]. Before the deposition of the wurtzite $Ti_{1-x}Al_xN$ films by arc deposition, the above chemical cleaning procedure was repeated for the TiN-coated substrates but only with acetone and ethanol.

*2.2 Cathodic Arc Deposition of w-$Ti_{1-x}Al_xN$*

The wurtzite phase $Ti_{1-x}Al_xN$ films were grown in an industrial-sized reactive cathodic arc system (KCS Europe AiPocket) equipped with a pulsed substrate bias power supply. Powder metallurgical Ti/Al cathodes (100 mm in diameter) from Plansee GmbH were used to grow coatings with different Al compositions. The nominal compositions of the cathodes were (Ti/Al) 33/67 at.%, 25/75 at.%, 15/85 at.%, and 5/95 at.%. The films were grown in a pure 6 Pa $N_2$ atmosphere at ~500°C. The substrate bias voltage was pulsed between ground and -50 V peak at a frequency of 10 kHz with a 10% duty cycle. The 3-minute deposition time and 2-fold substrate stage rotation resulted in a film thickness of ~300-500 nm, except for the 5/95 at.% cathode which suffered from bias voltage instabilities that resulted to a thickness of ~150 nm. More details on the growth and effects of duty cycle of the pulsed substrate bias voltage on the cathodic arc deposition process are detailed in Ref. [27].

*2.3 Annealing Experiments*

The arc-deposited wurtzite films were cut into 2.5 x 5 mm pieces and were then isothermally annealed at 950°C, 1100°C, or 1200°C using a tube furnace with a base pressure of ~5 ×$10^{-4}$ Pa. The films were heated at a rate of 20°C/min and the temperatures were maintained for 120 minutes. After annealing, the films were cooled to room temperature at a rate of 50°C/min.

*2.4 Film Characterization*

The compositions of the thin films were measured by time-of-flight elastic recoil detection analysis (ToF-ERDA) performed at the 5 MV 15SDH-2 tandem accelerator at the Tandem Laboratory, at Uppsala University. These measurements were carried out using a 36 MeV $^{127}I^{8+}$ beam at an incidence angle of 67.5° and a recoil angle of 45° with respect to the sample surface normal. The elemental depth profiles were analyzed with the Potku 2.0 software package [37].

X-ray diffractograms (XRD) of the thin films were measured using a PANalytical Empyrean diffractometer with Cu-K$\alpha$ radiation. Diffractograms for phase analysis were recorded with a channel-cut 2-bounce Ge(220) monochromator as primary optics. Pole figure measurements were obtained using a combination of x-ray lens as primary optics and parallel plate as secondary optics in a point focus configuration.

Morphological and cross-sectional imaging were done with a Zeiss Sigma 300 scanning electron microscope (SEM). Atomic-resolution microstructural information was obtained with an aberration-corrected FEI Titan³ scanning transmission electron microscope (STEM) operated at 300 kV. Three detectors were simultaneously used during STEM imaging: one bright field (BF), and two high-angle annular dark field (HAADF) detectors (FEI and Gatan). Compositional maps were obtained from an attached Bruker SuperX Energy Dispersive X-ray (EDX) detector. The TEM lamellae were prepared by an in-situ lift-out technique in an FEI Helios Nanolab 650 focused ion beam (FIB)-SEM dual beam system. Analyses and autocorrelations of the micrographs were carried out in Gatan's Digital Micrograph software.

Hardness and elastic modulus were determined by nanoindentation performed with a Hysitron TI950 TriboIndenter using a calibrated Berkovich diamond tip. Indentation was performed at fixed load (8-12 mN), which was low enough to avoid influence of the substrate. The hardness and elastic modulus were extracted from the unloading curve using the Oliver and Pharr method [38]. The values reported are averages from at least 25 indents for each sample with the corresponding standard deviations.

*2.5 Theoretical Calculations*

First-principles simulations were performed for the different $Ti_{1-x}Al_xN$ phases in the framework of the density functional theory (DFT) by the Vienna ab-initio simulation package (VASP) [39], using the projector augmented wave method and Perdew-Burke-Ernzerhof Generalized Gradient Approximation (PBE-GGA) [40]. Calculations were carried out for a range of Al-concentrations considering the cubic B1 and wurtzite B4 structures of $Ti_{1-x}Al_xN$. To model the B1 and B4 random alloys, special quasi-random structures (SQS) [41-43] were constructed with 128 atoms per supercell and the short-range order parameters optimized for pair interactions up to the 8$^{th}$ pair. Relaxation of each cell was done using a 5x5x5 mesh in the Brillouin zone and an energy cut-off of 600 eV was used.

The elastic tensor of TiAlN random alloys with cubic or hexagonal symmetry were calculated according to the procedure outlined by Tasnádi *et al.* [44]. It includes calculating the symmetry average of nine elements of the full elastic tensor ($C_{11}$, $C_{22}$,...). The results are utilized to determine the Voigt-Reuss-Hill averaged polycrystalline elastic moduli, such as bulk ($K_{VRH}$), shear ($G_{VRH}$), Poisson's ratio and Young's modulus E, according to [45]

$$E = \frac{9K_{VRH}G_{VRH}}{3K_{VRH} + G_{VRH}} \quad (1)$$

The directional elastic modulus was calculated from the elastic tensors based on the method by Ting [46]. In addition, the universal anisotropy of all the alloys was calculated according to [44]:

$$A^U = 5\frac{G_V}{G_R} + \frac{K_V}{K_R} - 6. \quad (2)$$

Here, the *V* and *R* indicate Voigt and Reuss polycrystalline elastic moduli, respectively. The universal anisotropy takes values of $A^U \geq 0$, where zero indicates an isotropic material.

The free energy of mixing $\Delta G$ for both cubic and wurtzite alloys were constructed from the DFT results according to:

$$\Delta G(x, p, T) = G(x, p, T) - xG_{AlN}(p, T) - (1-x)G_{TiN}(p, T) \quad (3)$$

where c-TiN and w-AlN were used as the reference phases, and the Gibbs free energy is a function of the fraction of Al in the alloy *x*, pressure *p* and temperature *T*. The configurational entropy contribution $S_{conf}$ to the Gibbs free energy was calculated in the mean-field approximation according to $S_{conf}(x) = -k_B(x\ln(x) + (1-x)\ln(1-x))$ [47]. Vibrational contributions to the free energy were calculated according to the quasi-harmonic method described by Moruzzi *et al.* [48]. The Debye temperatures were calculated based on the average sound velocities of the alloy systems, and a full description of the procedure can be found in the supplemental material. Strictly speaking, the method to account for vibrational contributions is applicable to elastically isotropic materials. Therefore, we did not use it for alloy compositions with estimated elastic anisotropy $A^U$ higher than 0.3. Fortunately, this limitation did not affect the construction of the phase diagram.

The phase diagram was constructed based on polynomial fits to the free energies of the cubic and wurtzite $Ti_{1-x}Al_xN$ phases without configurational contributions to entropy, with the polynomial degree chosen separately for each considered phase to avoid overfitting. Configurational contributions to free energy as a function of temperature were instead treated directly in the above-mentioned mean-field approximation. Binodal lines were obtained by the common tangent construction, while the spinodal lines, under which the alloy spontaneously decomposes isostructurally, were constructed for compositions and temperatures where the second derivatives of the free energy were zero. Since no contributions from coherency strains or energy bound in the compositional gradient are included in the expression for free energy, the calculated lines represent the chemical spinodal, which overestimates the compositional range in which spinodal decomposition is possible. We will, however, refer to them as spinodal lines throughout this work.

## 3. Results

*3.1 Chemical Composition of the $Ti_{1-x}Al_xN$ Films*

The results of the compositional analysis performed by ToF-ERDA are summarized in Figure 1a. The ToF-ERDA depth profiles show only minor compositional changes through the film thickness. Note that the TiN seed layer was excluded in Fig. 1a. As expected, increasing the Al content in the cathode also increased the amount of Al in the films. The Ti:Al cathode ratio was almost replicated in the resulting nitride films. Since the N content ranged from about 0.96 to 1.03, the films were considered stoichiometric. Less than 5% of oxygen and hydrogen contamination were detected for the cathodic arc deposited films, which is expected considering that the films were grown in an industrial chamber [2].

Figure 1b shows the x-ray diffractograms of the $Ti_{1-x}Al_xN$ films. Reference lines are superimposed to indicate the peak positions of relevant binary nitride phases (c-TiN, w-AlN). For all the samples, the peak found at ~36.6° corresponds to the magnetron sputtered TiN(111) seed layer while the intense peak at 36.9° corresponds to the MgO(111) substrate. The sample with x = 0.65 shows a broad hump between 33° and 36° corresponding to a w-$Ti_{1-x}Al_xN$ phase, and a peak at ~37.7° corresponding to a c-$Ti_{1-x}Al_xN$ phase. At x = 0.75, only one broad hump is observed between 33° and 36° again corresponding to a w-$Ti_{1-x}Al_xN$ peak. At x = 0.85 and 0.95, shoulders are found in the range of ~35.5° to 36.5°. When w-$Ti_{1-x}Al_xN$(0002) is grown epitaxially on to the c-TiN(111) template, the 0002 and 111 reflections are closely spaced and usually overlaps [49]. The appearance of peaks with broad full width half max (FWHM) and shoulders is also expected since cathodic arc deposition introduces a high number of lattice defects (impurities, vacancies, and interstitials) compared to high quality epitaxial films grown by, for example, magnetron sputtering. We also observed the layering effects due to substrate rotations (shown in the subsequent micrographs) which also contributes to peak broadening. Furthermore, the deposition temperature used (500°C) is too low to enable bulk diffusion.

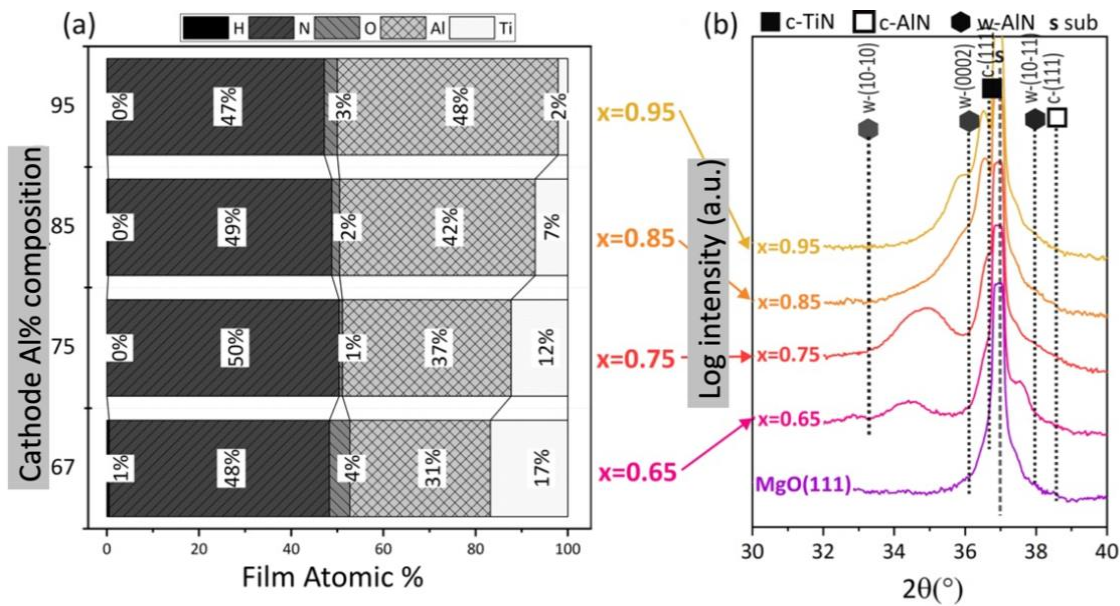

Figure 1. (a) Elemental composition of the $Ti_{1-x}Al_xN$ films as measured by ToF-ERDA. (b) XRD of the resulting film series with varying Al content. Similar scans were obtained for all the other samples, confirming the epitaxial growth on the TiN/MgO(111) substrate.

Figure 2 shows the azimuthal φ-scans of the w-$Ti_{1-x}Al_xN$ films with x=0.75, 0.85 and 0.95 and of the TiN/MgO(111) seed layer. The off axis 002 Bragg peaks of the c-TiN layer inclined 54° from the [111] direction shows a 3-fold symmetry, while the 10$\bar{1}$1 off axis peak of the wurtzite TiAlN film inclined ~60-62° from the w-[0001] direction exhibited a 6-fold symmetry. This confirms the pseudoepitaxial relationship of (111)$_{TiN}$ ∥ (0001)$_{w\text{-}TiAlN}$ and [110]$_{TiN}$ ∥ [11$\bar{2}$0]$_{w\text{-}TiAlN}$ between the film and the cubic TiN seed layer.

Figure 3 shows plan-view SEM images of the as deposited films. The film with x = 0.65 exhibits the roughest surface which is attributed to its polycrystalline nature containing both cubic and wurtzite phases.

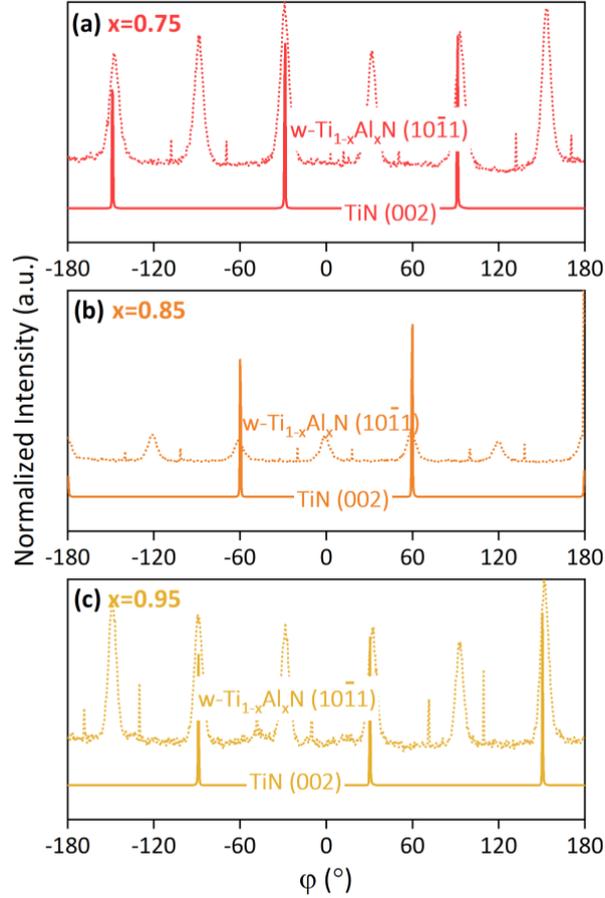

Figure 2. φ-scans of the w-Ti$_{1-x}$Al$_x$N films with (a) x=0.75, (b) x=0.85, (c) x=0.95 showing the w-10$\bar{1}$1 reflections compared to the c-002 reflections of the c-TiN /MgO(111) substrates.

Figure 4 shows the calculated phase diagram of wurtzite Ti$_{1-x}$Al$_x$N as a function of composition and temperature. The calculations show that there is a miscibility gap for a wide range of compositions with limited solubilities of Al in c-Ti$_{1-x}$Al$_x$N and Ti in w-Ti$_{1-x}$Al$_x$N. In addition, we estimate a tendency of the alloys towards a spontaneous isostructural decomposition by calculating the second derivative of the Gibbs free energy for both the B1 and B4 phases. When $\frac{\partial^2 G}{\partial x^2} < 0$, the solid solution is unstable with respect to concentration fluctuations and undergoes isostructural decomposition, provided that atoms have sufficient energy to diffuse. This spinodal decomposition is well known and well-studied for the cubic phase. The shape of the B1 spinodal line calculated for the cubic isostructural spontaneous decomposition in our case is in good agreement with earlier results by Alling *et al.* [28], see Supplementary Materials Fig. S2. Interestingly, our simulations predict spinodal decomposition also of the wurtzite phase, which would result in compositional modulations upon annealing of the wurtzite films. The wurtzite phase is mechanically unstable at T = 0 °K for Al contents less than 0.6875 since it fully disappears from the hexagonal Ti$_{1-x}$Al$_x$N energy landscape, which coincides with the composition at which cubic and wurtzite Ti$_{1-x}$Al$_x$N are energetically equally favorable in 0 °K DFT calculations. This instability makes it impossible to predict the wurtzite spinodal line for compositions with lower Al-content. In addition, we do not show high-temperature regions of the phase diagram because no liquid phases were included in our simulations. However, such high temperatures are not relevant for the interpretation of the experimental observations in this work.

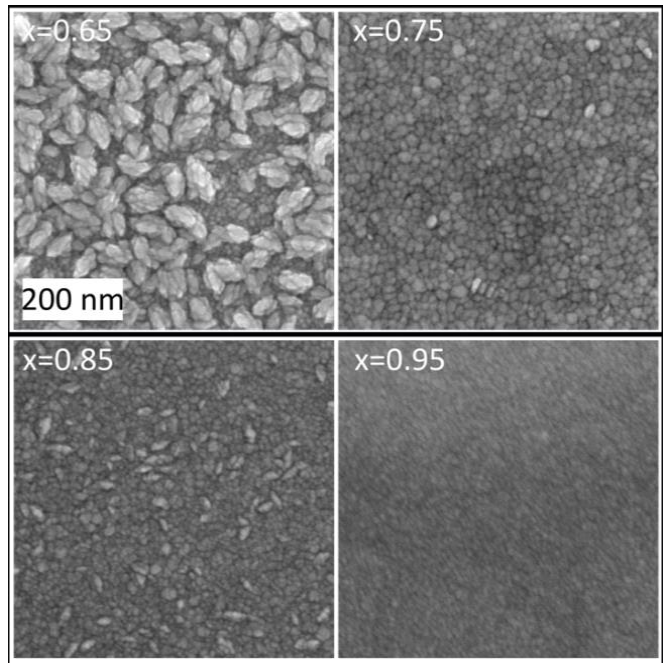

Figure 3. SEM plan view showing morphology of the arc-deposited $Ti_{1-x}Al_xN$ films.

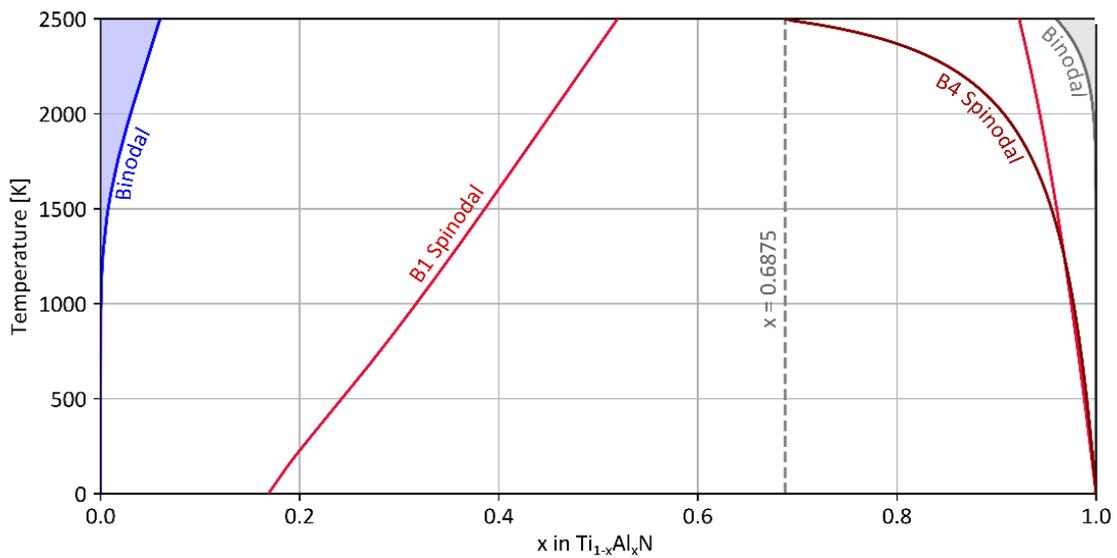

Figure 4. Calculated composition-dependent phase diagram of $Ti_{1-x}Al_xN$ alloys. Spinodal lines are drawn in light and dark red for cubic B1 $Ti_{1-x}Al_xN$ and wurtzite B4 $Ti_{1-x}Al_xN$, respectively, while binodal lines are shown as blue and gray. Shaded areas above the binodal lines indicate regions of stable cubic B1 $Ti_{1-x}Al_xN$ (shaded blue) in the Ti-rich and wurtzite B4 $Ti_{1-x}Al_xN$ (shaded gray) in the Al-rich regimes. The gray dashed line at $x = 0.6875$ denotes the limit of mechanical stability for w- $Ti_{1-x}Al_xN$ in 0 K DFT calculations.

Table 1. Calculated elastic tensor for B4 w-Ti$_{1-x}$Al$_x$N, along with calculated elastic moduli and Poisson ratio.

| Wurtzite (B4) Ti$_{(1-x)}$Al$_x$N | Elastic Tensor $C_{ij}$ [GPa] | | | | | | Elastic Moduli [GPa] | | | Poisson's ratio |
|---|---|---|---|---|---|---|---|---|---|---|
| | $C_{11}$ | $C_{22}$ | $C_{13}$ | $C_{12}$ | $C_{44}$ | $C_{66}$ | $K_{VRH}$ | $G_{VRH}$ | E | ν |
| Ti$_{0.3125}$Al$_{0.6875}$N | 325.5 | 139.3 | 140.9 | 158.4 | 107.9 | 83.6 | 162.4 | 71.8 | 187.8 | 0.307 |
| Ti$_{0.25}$Al$_{0.75}$N | 316.9 | 208.4 | 132.2 | 142.2 | 96.1 | 87.3 | 179.4 | 82.8 | 215.4 | 0.300 |
| Ti$_{0.1875}$Al$_{0.8125}$N | 338.4 | 261.5 | 117.7 | 137.7 | 102.2 | 100.3 | 185.2 | 98.2 | 250.4 | 0.275 |
| Ti$_{0.125}$Al$_{0.875}$N | 344.9 | 298.3 | 114.0 | 131.6 | 95.4 | 106.6 | 188.9 | 101.2 | 257.6 | 0.273 |
| Ti$_{0.0625}$Al$_{0.9375}$N | 360.4 | 316.0 | 107.7 | 131.1 | 108.4 | 114.7 | 191.4 | 112.2 | 281.6 | 0.255 |
| AlN | 376.2 | 355.3 | 98.1 | 128.5 | 122.5 | 123.9 | 194.4 | 121.6 | 302.0 | 0.242 |

Table 2. Calculated elastic tensor for well-known B1 c-Ti$_{1-x}$Al$_x$N, along with calculated elastic moduli and Poisson ratio.

| Cubic (B1) Ti$_{(1-x)}$Al$_x$N | Elastic Tensor $C_{ij}$ [GPa] | | | Elastic Moduli [GPa] | | | Poisson's ratio |
|---|---|---|---|---|---|---|---|
| | $C_{11}$ | $C_{12}$ | $C_{44}$ | $K_{VRH}$ | $G_{VRH}$ | E | ν |
| TiN | 613.6 | 109.9 | 181.4 | 277.8 | 206.9 | 504.9 | 0.202 |
| Ti$_{0.875}$Al$_{0.125}$N | 541.1 | 135.9 | 158.9 | 271.0 | 175.1 | 463.0 | 0.234 |
| Ti$_{0.75}$Al$_{0.25}$N | 512.5 | 142.1 | 175.6 | 265.5 | 179.4 | 449.3 | 0.224 |
| Ti$_{0.625}$Al$_{0.375}$N | 478.7 | 151.9 | 188.7 | 260.8 | 178.2 | 439.0 | 0.222 |
| Ti$_{0.5}$Al$_{0.5}$N | 463.7 | 152.9 | 207.1 | 256.5 | 184.6 | 438.5 | 0.210 |
| Ti$_{0.375}$Al$_{0.625}$N | 441.8 | 170.0 | 216.8 | 260.6 | 179.8 | 446.7 | 0.220 |
| Ti$_{0.3125}$Al$_{0.6875}$N | 429.9 | 162.7 | 222.8 | 251.7 | 181.5 | 435.3 | 0.209 |
| Ti$_{0.25}$Al$_{0.75}$N | 424.9 | 165.0 | 238.3 | 251.6 | 186.8 | 439.2 | 0.202 |
| Ti$_{0.125}$Al$_{0.875}$N | 416.2 | 168.8 | 262.4 | 251.3 | 194.0 | 432.3 | 0.193 |
| AlN | 424.9 | 167.8 | 306.0 | 253.5 | 216.1 | 497.3 | 0.168 |

Tables 1 and 2 show the full elastic tensors for wurtzite (B4) and cubic (B1) Ti$_{1-x}$Al$_x$N for different Al contents, respectively. It is worth highlighting that the calculated elastic tensors for the cubic Ti$_{1-x}$Al$_x$N phase (Table 2) using the methods described above agrees well with previously reported values [50,51], underscoring the validity of our calculations.

Figure 5a shows the calculated directional elastic modulus for B4 TiAlN with x = 0.75, 0.875, and 0.9375 which are closest to the compositions investigated experimentally. The 2D plots are cut-outs of the 3D directional images along the same directions as the high-res STEM measurements in Figure 7. In 5a, the axes only serve as a notion of direction, and actual values of the elastic modulus in any direction are given by the distance from the surface to the origin. Figures 5b and 5c show the elastic moduli of all structures of Ti$_{1-x}$Al$_x$N (at 0 K) and their calculated universal anisotropy, respectively. In Figure 5b, the elastic modulus of the B4 Ti$_{1-x}$Al$_x$N increases with increasing Al. The superimposed experimental elastic moduli obtained from the as-deposited wurtzite films follow this predicted increasing trend. In Figure 5c, a sharp increase in anisotropy can be noted in the B4 w-Ti$_{1-x}$Al$_x$N phase as the Al content is decreased toward x = 0.6875.

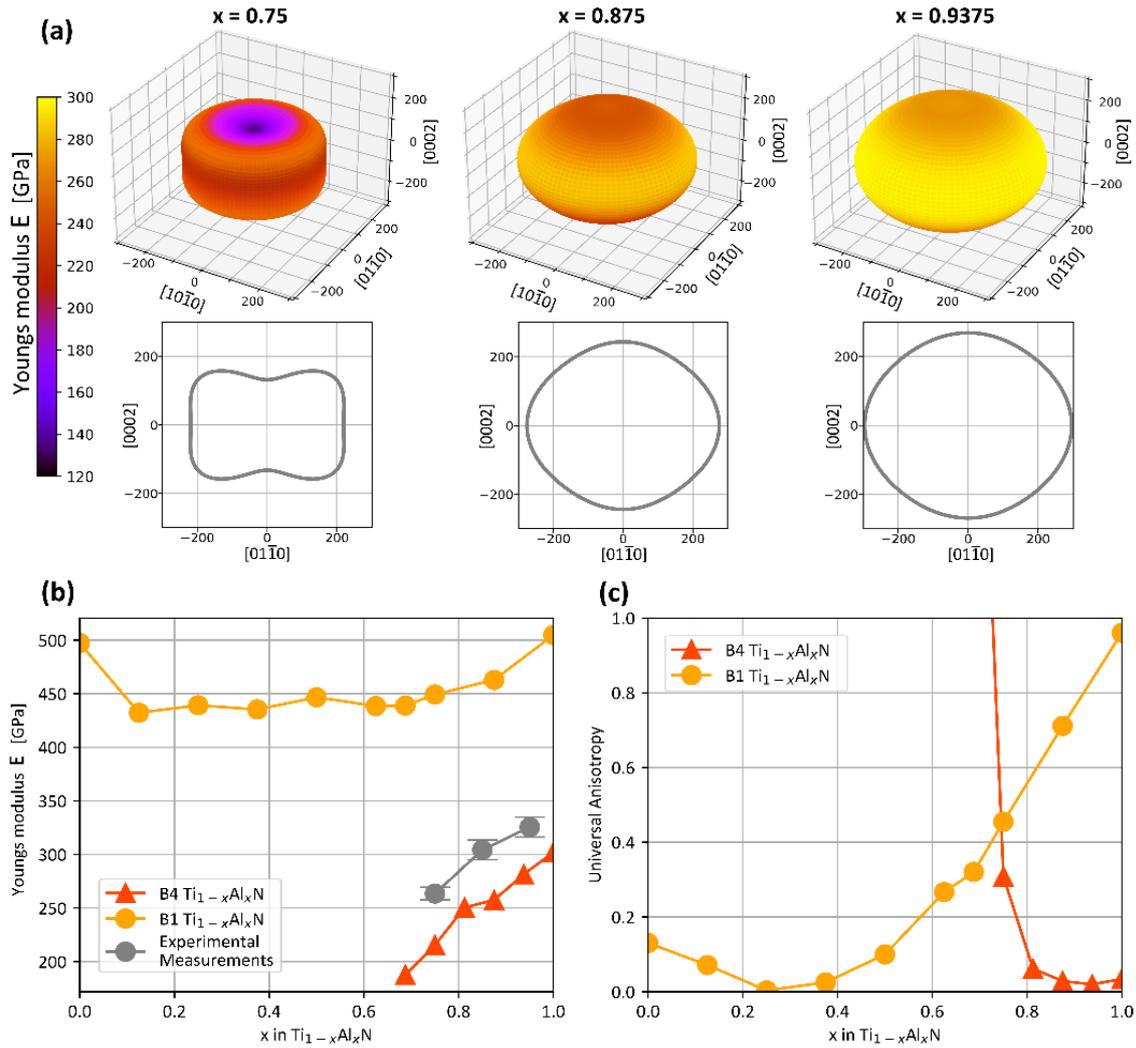

Figure 5. Calculated (a) directional elastic modulus (at 0 K) for B4 Ti$_{1-x}$Al$_x$N at x = 0.75, 0.875 and 0.9375, (b) elastic modulus E, and (c) universal anisotropy of the B1 and B4 phases of Ti$_{1-x}$Al$_x$N.

Figure 6 displays the x-ray diffractograms of the w-Ti$_{1-x}$Al$_x$N films after annealing at different temperatures. The structures of the films remain predominantly composed of wurtzite Ti$_{1-x}$Al$_x$N solid solutions, although slight changes in the wurtzite peak widths and shapes were observed. In addition, after annealing these samples to 950°C, we see formation of shoulders in the 2θ range of 37 to 38°. These shoulders are interpreted as formation of small amounts of a cubic solid solution Ti$_{1-x}$Al$_x$N phase containing high Al content.

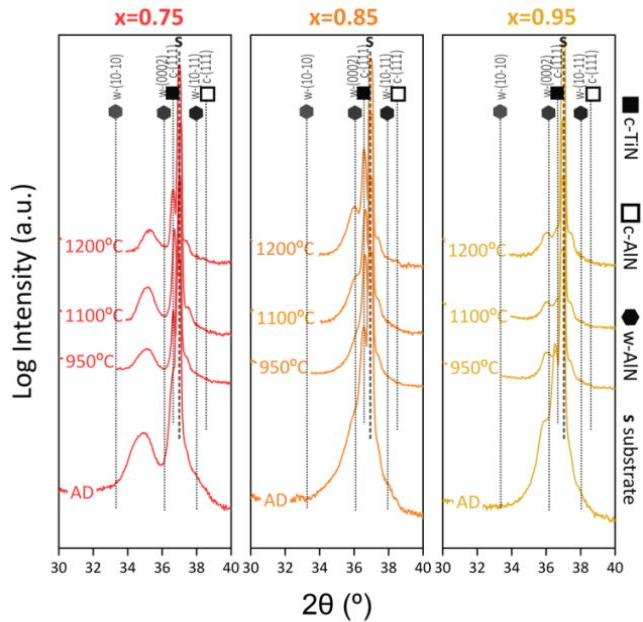

Figure 6. X-ray diffractograms of the w-Ti$_{1-x}$Al$_x$N with x = 0.75, 0.85 and 0.95 annealed at 950°C, 1100°C and 1200°C.

Figure 7 shows the respective cross-sectional HAADF-STEM micrographs, high resolution STEM, and EDX maps of the Ti$_{1-x}$Al$_x$N films with (a-c) x = 0.75. (d-f) x = 0.85, and (g-i) x = 0.95 after annealing for 2 hours at 950°C. The darker contrast regions are rich in Al while brighter contrast regions are rich in Ti in the HAADF-STEM images, which is also confirmed by the compositional EDX maps for all the three samples. The change in contrast for films with x = 0.75 and x = 0.85 indicates chemical segregation into Ti-rich and Al-rich regions.

At x = 0.75 (Fig. 7a,b), the images show compositional modulations similar to the ones observed in w-Zr$_{1-x}$Al$_x$N [52]. At higher resolution (Fig. 7b), the observed compositional domains are elongated in the <01$\bar{1}$0> direction and appear throughout the film thickness. The FFT analysis shown in the inset in Fig. 7b shows that the sample retains its wurtzite structure despite the chemical decomposition, although some arcing of the FFT spots is noted.

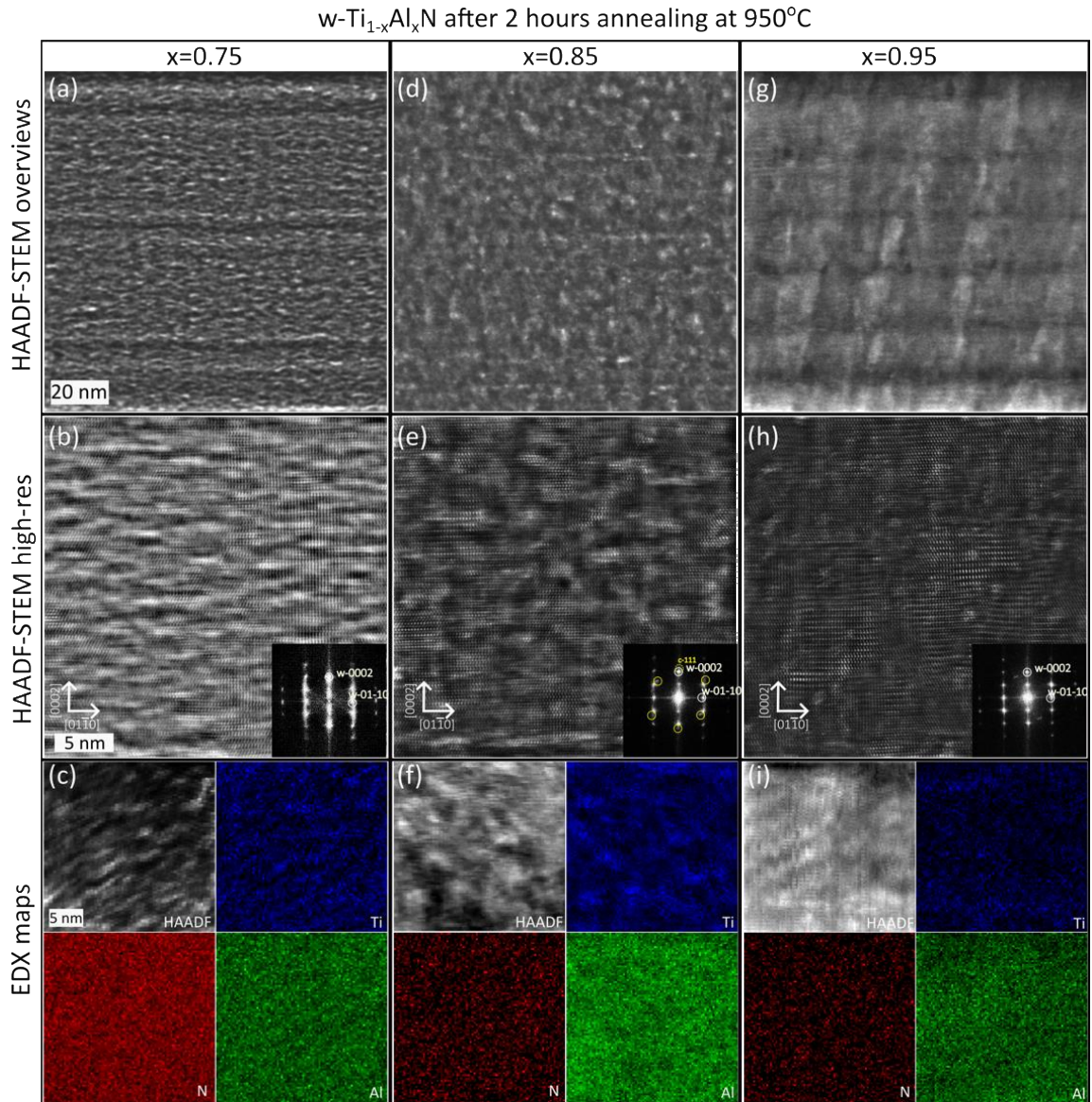

Figure 7. Cross-sectional HAADF-STEM (a-b, d-e, g-h) and EDX composition (c, f, i) maps of the wurtzite $Ti_{1-x}Al_xN$ films with x = 0.75, 0.85, 0.95 after annealing for 2 hours at 950°C. Images are scaled in the same way for easier comparison between samples.

For the w-$Ti_{1-x}Al_xN$ film with x = 0.85, chemical segregation can also be observed, confirmed by EDX, however it is less pronounced, with less periodic compositional modulations. Instead, the TiN-rich domains (bright contrast regions) are non-uniform in size and more randomly distributed. The corresponding FFTs (insets in Fig. 7e) confirm the predominant wurtzite structure, as also observed in XRD. However, new FFT reflections indicated by yellow circles in Fig. 6e have also developed, which indicates the formation of a cubic phase, i.e. w-$Ti_{0.15}Al_{0.85}N$ (0001)[11$\bar{2}$0] || c-TiN (111)[110], similar to what was observed by Calamba et al. [25].

It is uncertain whether chemical decompositions can be observed in the w-$Ti_{1-x}Al_xN$ film with x = 0.95 (as shown in Figure 7g). The concentration of Ti in the film (~5%) is close to the detection resolution limit of EDX when it is initially deposited, making it difficult to determine whether segregation is occurring. The FFT inset in Figure 7h shows only wurtzite structure of this film after annealing to 950°C.

Figure 8 shows the microstructure evolution of w-$Ti_{0.25}Al_{0.75}N$ (a-c) and w-$Ti_{0.15}Al_{0.85}N$ (d-f) after annealing for 2 hours at 1100°C. High-resolution STEM (Fig. 8b,e) and the corresponding inverse FFT lattice fringes of the 0002 spots are shown in Fig. 8c and 8f for each film. Dislocations and stacking faults were identified in the inverse FFT images and translated to the high-resolution STEM images.

For w-Ti$_{0.25}$Al$_{0.75}$N (x = 0.75), compositionally elongated modulations are still observed. We note that the Ti-rich regions (bright contrast) still cluster in layers of two lattice planes and maintain coherency with the hexagonal lattice structure. The corresponding FFT (inset in Fig. 8a) also shows more arcing of the FFT. Meanwhile for w-Ti$_{0.15}$Al$_{0.85}$N (x = 0.85), the decomposing domains (exhibited by the dark and bright contrast regions) are slightly larger in size than at 950°C. Still, they are randomly dispersed, and their evolving shape is non-uniform. The inset FFT in Fig. 8e clearly shows the evolution of new faint spots, encircled in yellow, again indicating the formation of cubic domains. They maintain a semi coherent relationship with the surrounding wurtzite domains.

To quantitatively estimate the compositional fluctuations observed in the annealed w-Ti$_{0.25}$Al$_{0.75}$N films, we obtained the autocorrelation functions of the HAADF-STEM images. Figure 9 shows the plots of the lineouts of the autocorrelated image. The lineouts were extracted in both the [0001] growth direction and the [01$\bar{1}$0] in-plane direction. The autocorrelations of w-Ti$_{0.25}$Al$_{0.75}$N after annealing to 950°C and 1100°C show oscillations with periods of ~2.5-3 nm along the [0001] direction. The lineouts along the [01$\bar{1}$0] direction however do not show distinct periodicity in the oscillations. (Note that the smaller ~0.25 nm oscillations are due to the lattice-resolved intensity changes.) The w-Ti$_{0.15}$Al$_{0.85}$N (x = 0.85) film does not display a distinct oscillation period for any direction or temperature.

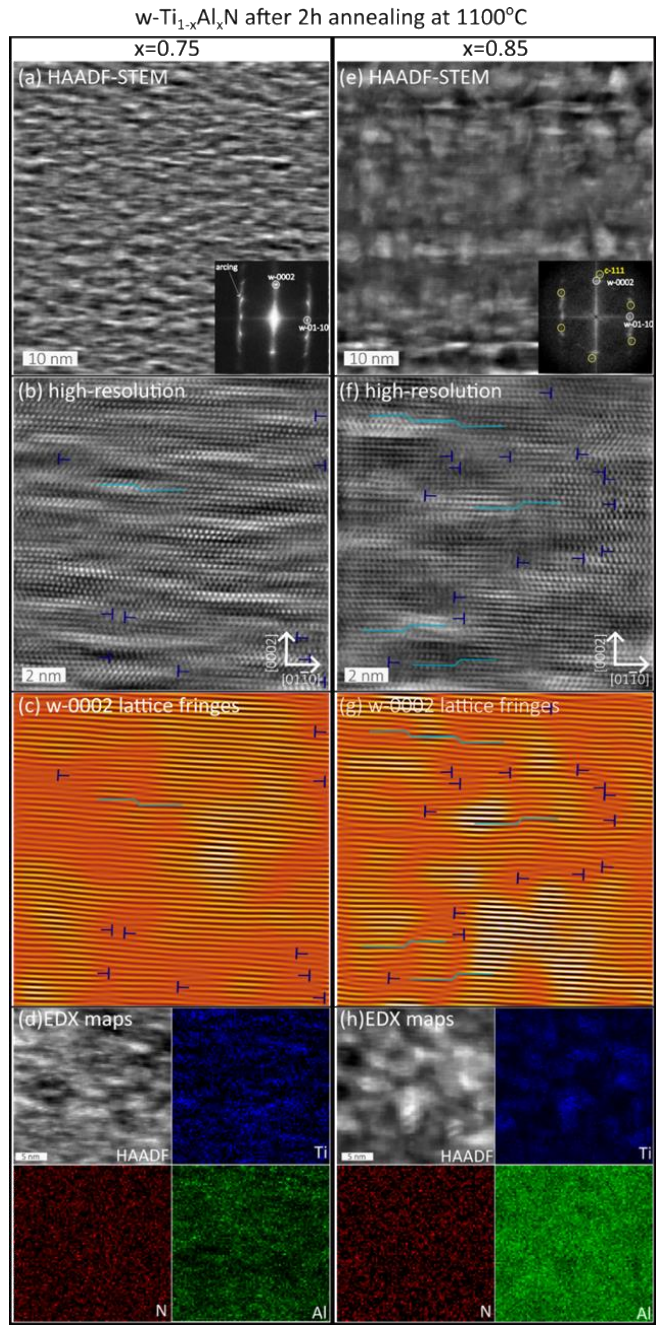

Figure 8. High-resolution HAADF-STEM images, EDX maps, and inverse FFTs of the w-Ti$_{1-x}$Al$_x$N samples with (a-d) x = 0.75 and (e-h) x = 0.85 after 2 hours of annealing at 1100°C.

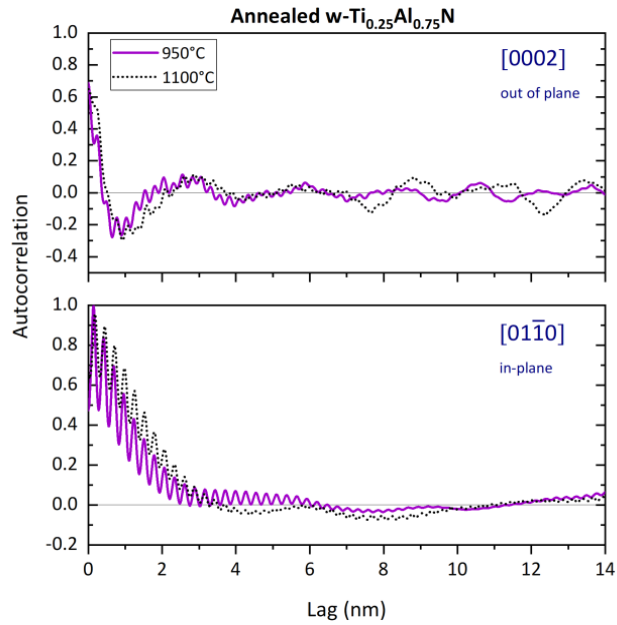

Figure 9. Autocorrelation function lineouts obtained from the HAADF-STEM images of w-Ti$_{0.25}$Al$_{0.75}$N films annealed at 950°C (solid line) and 1100°C (dotted line). The lineouts were extracted in the (a) [0002] growth direction and (b) [01$\bar{1}$0] in-plane direction.

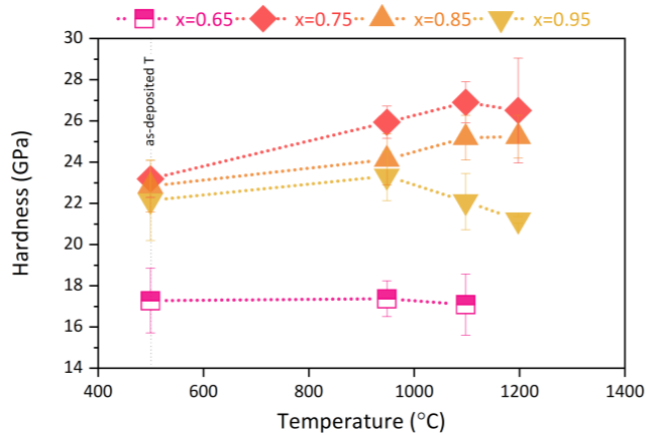

Figure 10. Hardness of the Ti$_{1-x}$Al$_x$N films obtained by nanoindentation plotted versus annealing temperatures.

Figure 10 shows the hardness measured by nanoindentation after annealing to 950°C, 1100°C, and 1200°C. In the as-deposited state, the hardness of the w-Ti$_{1-x}$Al$_x$N samples with x = 0.75, 0.85, 0.95 fall in the range of 22 to 23 GPa. The sample with x = 0.65, exhibiting a dual phase mixture of cubic and wurtzite already in its as-deposited state, has a lower hardness (~17 GPa). Annealing to higher temperatures did not significantly change its hardness either. The low hardness of the film with x = 0.65 is attributed to its polycrystalline nature with incoherent grain boundaries that already formed during growth [7], in combination with the known under-dense structure for cubic films grown with a low duty cycle of the pulsed substrate bias voltage [27]. After annealing to 950°C, the hardness of the samples with x = 0.75, 0.85, and 0.95 increased by ~1-2 GPa, which suggests age hardening of the pure wurtzite phases. Both w-Ti$_{0.25}$Al$_{0.75}$N and w-Ti$_{0.15}$Al$_{0.85}$N continued to increase in hardness up to an annealing temperature of 1100°C. At 1200°C the film with x = 0.75 showed a slight decrease in hardness while the film with x = 0.85 retained its hardness. The film with x = 0.95 started to decrease in hardness after annealing to 1100°C.

## 4. Discussion

*4.1 Growth of metastable single-phase wurtzite solid solution of $Ti_{1-x}Al_xN$*

Our combined experimental and theoretical approach shows that metastable wurtzite TiAlN with $x \geq 0.75$ can be grown by PVD, particularly cathodic arc deposition. This allows for a fundamental study of single-phase wurtzite TiAlN without contributions from a secondary cubic TiAlN phase (*e.g.* interdiffusion, structural, and strain effects). In our current work, we have shown that by using a low-duty cycle pulsed substrate bias technique to cathodic arc deposition and using high Al-containing cathodes in a pure nitrogen environment, we could promote epitaxial growth on a suitable template. By this technique, we have combined the effects of high deposition rates and low average ion energies [27] to promote a metastable and local pseudo-epitaxial growth while maintaining the aluminum to titanium ratio of the cathode in the film. The combined use of a single crystalline substrate and these kinetically limited growth conditions limit the mobility of adatoms, which lead to growth of closed packed surfaces [53]. In this case, the (111) orientation of TiN was chosen as a seed layer for the growth of wurtzite $Ti_{1-x}Al_xN$ because it has been shown to promote growth of wurtzite domains along the c-axis, *i.e.* (0001) [26].

We also observed a single orientation of the wurtzite crystallites (0001) on the samples with $x = 0.75$ to $0.95$ and did not exhibit any chemical segregation in their as-deposited states (confirmed by STEM-EDX, not shown). Previous studies of the growth of wurtzite $Ti_{1-x}Al_xN$ using high Al containing (>50%) TiAl alloy targets have always entailed the growth of a cubic $Ti_{1-x}Al_xN$ layer first [25,26], followed by the stabilization of wurtzite layer. The formation of the wurtzite layer after a critical thickness was attributed to the increased strain and stacking faults which help facilitate the cubic-to-wurtzite layer transition [25,54]. This consequently affects the stability of the wurtzite phase in its as-deposited form, wherein nanoscale segregations of Ti-rich and Al-rich regions were already observed at the growth front [25] but not in the cubic $Ti_{1-x}Al_xN$ layer below. Thus, the main difference of the technique used herein is that we can grow undecomposed, single-phase, and homogenous wurtzite $Ti_{1-x}Al_xN$ solid solutions without the co-growth of cubic $Ti_{1-x}Al_xN$ layers or grains.

However, as cathodic arc deposition remains more energetic than DC or RF magnetron sputtering techniques, it is likely that it produces films with significantly higher defect densities. These defects can be in the form of self-interstitials, anti-sites, and metal vacancies, as well as local disruptions in the lattice such as dislocations and stacking faults that were generated by the collision cascade. The combination of these defects affect the lattice parameters and peak widths measured by XRD [49], which is why we obtained broad peaks and shoulders.

*4.2 Elastic Constants*

We present the entire elastic tensor as a function of x for the wurtzite B4 $Ti_{1-x}Al_xN$ alloy. As expected, wurtzite $Ti_{1-x}Al_xN$ is more compliant than the cubic alloys at any Al-concentration. The universal anisotropy factor shows that as the aluminum content in w-$Ti_{1-x}Al_xN$ increases, the material becomes increasingly isotropic, as expected for pure AlN [55]. However, anisotropy sharply increases as the aluminum content approaches $x \approx 0.6875$. The cubic $Ti_{1-x}Al_xN$, anisotropy values agree well with the previously published observation of isotropy at around $x = 0.3$ Initially, the layered hexagonal Bk $Ti_{1-x}Al_xN$ phase was also considered but it was excluded due to considerable anisotropy throughout its entire stability range as detailed in the supplemental material. Experimental measurements of elastic modulus E in the wurtzite $Ti_{1-x}Al_xN$ films in Figure 5b follow the same trend as the theoretically calculated values.

*4.3 Microstructure Evolution During the Decomposition of Wurtzite $Ti_{1-x}Al_xN$*

The combined theoretical and experimental results suggest the existence of a spinodal region and a significant driving force for spontaneous decomposition in the high-Al regime of the wurtzite $Ti_{1-x}Al_xN$ phase. This decomposition was confirmed by high-resolution Z-contrast STEM images and EDX analyses wherein the annealed wurtzite films clearly show a decomposed structure, as exhibited by the dark and bright contrast variations. However, the mechanism of decomposition remains to be discussed.

For the w-$Ti_{0.25}Al_{0.75}N$ film (Fig. 7a-c), discernible nanoscale compositional fluctuations similar to the decomposed microstructure of wurtzite $Zr_{1-x}Al_xN$ coatings were observed [52]. At the same time, the wurtzite 0002 diffraction peak remains, although its FWHM is slightly changing. Previous studies assumed that the spinodal decomposition of alloys with wurtzite structure occurs through the formation of layers parallel to the (0001) plane [56], similar to what is observed for our sample with x = 0.75. However, it was later determined that there are other factors influencing this preferential orientation of the decomposed layers, such as composition, surface energies [57], interfacial energies [58,59], coherency strains [56], and stress of the coating where the film can be elastically strained [60]. For example, it was calculated for CrN [58], TiN, and ZrN [61], that the interfacial energies between their cubic and AlN wurtzite structures are lower when the interface follows the relation c-(111)∥w-(0001) compared to other possible semicoherent interfaces. As this Al composition (x = 0.75) is close to the composition where wurtzite and cubic have the same free energy [32], we infer that the two atomic layer thin Ti-rich domains have phase transformed to the cubic phase by a mechanism mediated by partial dislocations causing (111) stacking faults [62]. The cubic phase sustains a coherent relationship with the surrounding hexagonal matrix through $(111)_{c\text{-TiN}}\|(0001)_{w\text{-AlN}}$ interfaces.

We note that the autocorrelation of the chemical fluctuation shows that there is periodicity along [0002] but not along [01$\bar{1}$0] direction, which suggests strong crystal templating influences on the segregation. The elastic anisotropy differences of the evolving domains, while they are known for cubic $Ti_{1-x}Al_xN$ [50,63], are yet to be discussed for wurtzite $Ti_{1-x}Al_xN$. Our calculations in Figure 5a suggest that w-$Ti_{1-x}Al_xN$ with x = 0.75 has significantly higher anisotropy than those with higher x values which could be influencing the decomposition and microstructure evolution.

With a ~10% increase in Al content (*i.e.,* x = 0.75 to 0.85), we observed through high-resolution HAADF-STEM imaging that the decomposition behavior drastically changed. The decomposition of the w-$Ti_{0.15}Al_{0.85}N$ film shows compositional fluctuations, but the sizes of the domains are larger and less uniform than the ones observed for x = 0.75. Two possible explanations are suggested. One, considering that the sample was annealed for 2 hours at 950°C, the sample could well be in the stage of coarsening following spinodal decomposition. Second, the film could be decomposing in a non-ideal way, which can happen to compositions close to the spinodal line [64]. Comparing the two samples, x = 0.75 and 0.85, the modulation size also decreases with increasing Ti content, similar to the trends w-ZrAlN coatings with increasing Zr [52]. This observation also follows the expected trend wherein the wavelength of the chemical fluctuation scales inversely with driving force for decomposition [2,65]. In this case, the sample with x = 0.75 shows smaller compositional wavelength, which indicates a higher driving force for decomposition.

Meanwhile it is unclear whether the sample with x = 0.95 exhibits decomposition, as the amount of the secondary Ti-rich domains that could form is relatively small and proven difficult to detect.

*4.4 Age Hardening*

The observed age hardening (see Fig. 10) of the wurtzite films after annealing to high temperatures can be attributed to the microstructure resulting from the decomposition of the films. The coherency strains between the evolving wurtzite Ti-rich domains and Al-rich domains, despite their differences in the decomposition routes, induced hardening. Specifically, the binary w-AlN phase has an out-of-plane lattice parameter of c = 4.98 Å and an in-plane lattice parameter of a = 3.11 Å [66], which corresponds to differences with the w-$Ti_{1-x}Al_xN$ of up to 3%, depending on the Ti to Al ratio. The different domains have different elastic properties, which adds to hardening. In addition, annealed microstructures will also lead to different macroscopic mechanical properties, *i.e.* elongated domains strengthen the material more than spherical domains [67].

Further, high-resolution lattice fringe analyses show the presence of stacking faults and extra (0002) planes along with misfit dislocations (Fig. 8) which form semi coherent interfaces in some regions of the film. These evolving interfaces, which are both coherent and semi coherent, offer the benefits of coherency as well as Koehler hardening [50,68]. This semi coherency continued when the films were annealed up to 1100°C, as shown by the widening in the XRD peaks and high-resolution STEM (Fig. 8) and explains the continued hardness increase.

After annealing at 1200°C for 2 hours, the coherency is reduced and point defects are annihilated which resulted in a decrease in hardness (over aged). Further, at higher temperatures, the growth and coalescence of domains likely cause a breakdown of the semi coherency in order to reduce strain energy.

Overall, these findings show that for a narrow high Al composition range, the pure wurtzite phase can still contribute to age hardening, especially in high temperature applications where decomposition is initiated. For comparison, the mixed cubic and wurtzite phase (film with x=0.65) did not exhibit hardening and it is attributed to the as-grown incoherent interfaces between the cubic and wurtzite phases.

## 5. Conclusion

The phase stability, decomposition, and mechanical properties of single-phase wurtzite $Ti_{1-x}Al_xN$ (x = 0.65, 0.75, 0.85, 0.95) films synthesized by low-duty cycle pulsed substrate bias cathodic arc deposition were investigated experimentally and theoretically. Films with x ≥ 0.75 exhibited local epitaxial growth while the film with x = 0.65 showed a mixed fiber texture with secondary cubic grains. First-principles DFT calculations predicted the elastic moduli to decrease and the elastic anisotropy to increase with increasing Ti-content of the wurtzite $Ti_{1-x}Al_xN$ phase. Phase diagram simulations demonstrate the existence of a spinodal region, which is in accord with the decomposition paths experimentally observed. High-resolution HAADF-STEM and EDX mapping after vacuum annealing displayed compositional variations that change with increasing Al content. The film with x = 0.75 exhibited more elongated compositional modulations while the film with x = 0.85 exhibited a non-uniform and less elongated modulation after annealing to high temperatures. These chemical modulations resulted to age hardening with a hardness increase of ~1-2 GPa after an anneal at 950°C.

**Acknowledgements**

Theoretical modelling and calculations were carried out using the resources provided by the Swedish National Infrastructure for Computing (SNIC) and the National Academic Infrastructure for Supercomputing in Sweden (NAISS), partially funded by the Swedish Research Council through Grant Agreement no VR-2015-04630. We acknowledge the financial support given by the following: VINNOVA (FunMat-II project grant no. 2016-05156), the Swedish Research Council (VR grants no 2017-03813, 2017-06701, 2021-04426, and 2021-00357) and grant no 2019-00191 (for accelerator-based ion-technological center in Tandem accelerator laboratory in Uppsala University), and the Swedish government strategic research area grant AFM – SFO MatLiU (2009-00971). IAA and FT acknowledge support from the Knut and Alice Wallenberg Foundation (Wallenberg Scholar grant no. KAW-2018.0194).